# Preventing Distributed Denial-of-Service Attacks on the IMS Emergency Services Support through Adaptive Firewall Pinholing

Andreea Ancuta Onofrei, Yacine Rebahi, Thomas Magedanz
Fokus Fraunhofer Institute, Berlin, Germany
{andreea.ancuta.onofrei,yacine.rebahi,thomas.magedanz}
@fokus.fraunhofer.de

February 10, 2010

### Abstract

Emergency services are vital services that Next Generation Networks (NGNs) have to provide. As the IP Multimedia Subsystem (IMS) is in the heart of NGNs, 3GPP has carried the burden of specifying a standardized IMS-based emergency services framework. Unfortunately, like any other IP-based standards, the IMS-based emergency service framework is prone to Distributed Denial of Service (DDoS) attacks. We propose in this work, a simple but efficient solution that can prevent certain types of such attacks by creating firewall pinholes that regular clients will surely be able to pass in contrast to the attackers clients. Our solution was implemented, tested in an appropriate testbed, and its efficiency was proven.

## 1 Introduction

The transition to Next Generation Networks (NGNs) is often coupled with the vision of innovative services providing personalized and customizable services over an all-IP infrastructure. To enable a smooth transition, all the next generation IP networks need not only to support more services but also current vital services such as emergency services. The latter simply means placing calls to police, ambulance and fire brigade to ask for assistance or to inform about an incident.

The IP Multimedia Subsystem (IMS) [2] is in the heart of the NGNs technologies and is certainly the future replacement of the current telecommunication networks. As a consequence, 3GPP [3] has carried the burden of specifying a standardized IMS-based emergency services framework as it will be discussed later on.

Like any other communications network, emergency communications will be for sure the target of misuse and attacks. However, there are some special characteristics of the emergency services that allow the emergence of new kinds of attacks that are not visible in other communications networks. These characteristics are:

- Emergency systems are special-purpose networks with asymmetric network behavior: in case of a high-impact emergency event, the system will have to process a huge number of requests, while at most other times there will be very few requests.

- Emergency services are a prioritized service. Emergency messages have an emergency indication attached that guarantees to the messages to be transported immediately, and should not be delayed by other traffic [17].

- Emergency systems need to be available all time with few obstacles to the access. This means that all the relevant information should be gathered easily. On the other hand, the system should be prevented from misuse and attacks, i.e. the user might need to be authenticated. These two goals are mutually exclusive and one has to find a trade off between them.

Session Initiation Protocol [1] has established itself as the de-facto standard for VoIP services in the Internet and is the basis and key protocol of currently defined Next Generation Networks, like the IP Multimedia Subsystem (IMS). DoS attacks against the main components of the IMS-based emergency framework are a serious threat. Through excessive message flooding, either from regular accounts or generated by distributed attack tools, such attacks aim to render the emergency service inoperable by exceeding the components' message processing capability. Especially when the attack is conducted from multiple different sources (using so-called bot-nets) with seemingly valid and conforming SIP emergency messages, defence is a challenging problem, as the target Emergency component needs to distinguish between valid SIP emergency messages from regular users and DoS messages from malicious users. Such messages need to be dropped beforehand to keep the service operational.

The aim of this paper is to introduce a lightweight security mechanism based on firewall pinholing, that effectively prevents many DDoS attacks on the IMS based emergency framework.

Pinholing is a common firewall technique, where the configuration of the border firewall of the protected network is dynamically updated depending on current network traffic. The firewall is initially configured to block most incoming traffic, but allows some exceptions ("pinholes"), so that traffic with special characteristics can pass the firewall barrier. The proposed mechanism controls a firewall to generate pinholes that are necessary to effectively protect the emergency framework. The approach shares some similarities with greylisting [15] used in email spam prevention, i.e. all incoming requests are initially held back by the firewall (they are "greylisted"), and only forwarded to the destination if the sending entity follows the SIP emergency specification correctly. Hence, with our mechanism we can deny access to all distributed flooding bots that do not meticulously follow the specification. The mechanism cannot handle all types of DDoS attacks, however it is especially effective against flooding bots that utilise spoofed IP addresses, a common technique that attackers use to evade detection. Spoofed addresses are difficult to handle with all current prevention mechanisms.

We will introduce basic concepts later in the document, including IMS, emergency support for IMS, DoS, firewalls and greylisting. We will examine our work in the context of current state of the art. We will then explain our mechanism in detail and show performance measurements within our prototype security solution and test bed. We will conclude with possibilities for optimisation and will give hints about further work.

## 2 Background Information

### 2.1 IMS Overview

The IP Multimedia Subsystem (IMS) is the key enabler in the mobile world for providing rich multimedia services to the end-users. IMS first appeared in 3GPP release 5 of the evolution from 2G to 3G networks "from W-CDMA to UMTS". This release supports both GSM and GPRS networks. In 3GPP release 6, interworking with WLAN was added. 3GPP release 7 adds support for fixed networks, together with TISPAN [4], this collaboration allowed the adoption of a more generalized model able to address a wider variety of network and service requirements.

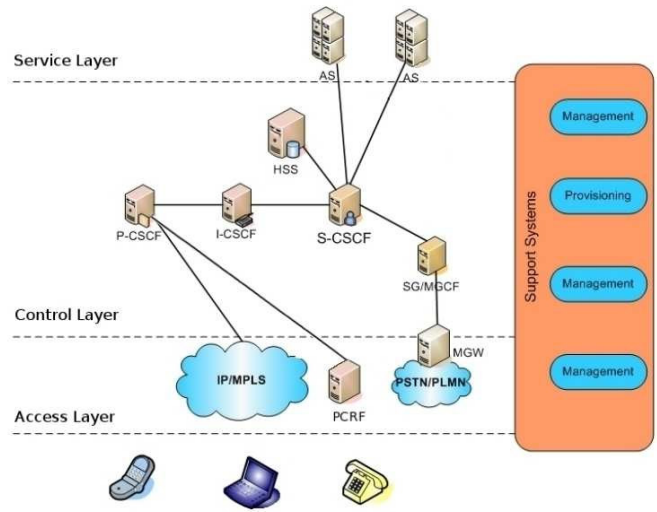

Figure 1: IMS architecture

The IMS underlying network architecture can be divided into three main layers: Access Layer, Control Layer and Service Layer (see Figure 1).

- The access layer consists of IP routers and legacy PSTN switches that provide access to the IMS network both from contemporary IP telephony devices and older circuit switch devices respectively. IP devices compatible with IMS incorporate a SIP user agent that is used to place voice or video calls toward the network.

- The control layer of the IMS network consists of nodes for managing call establishment, management, and release, namely the Call Session Control Function (CSCF), performed by: Proxy (P-CSCF), Interrogating (I-CSCF) and Serving (S-CSCF). Before the user can use the services from the IMS network it must authenticate itself, by performing a registration. The subscriber data of every user is located in the Home Subscriber Service (HSS), which acts as a Authentication, Authorization and Accounting (AAA) server, providing a central repository of user-related information. The P-CSCF is the link of the User Element (UE) to the IMS network, receiving all the signaling traffic from/for the user, allowing access only to registered users. The S-CSCF performs routing traffic towards other networks, manages billing and session expiration intervals, and interrogates the HSS to retrieve authorization, service triggering information and user profile. The I-CSCF is in charge of querying the HSS if a specific user is present at the HSS and which S-CSCF the HSS has allocated for it.

- IMS applications are hosted in the service layer. This layer consists of SIP Application Servers (AS) which provides the end user service logic. The (AS) execute IMS applications and services by manipulating SIP signaling and interfacing with other systems. Usually, the AS will offer a programming language and framework for creating new services, for example Java SIP and HTTP Servlets.

## 2.2   Emergency Services Support for IMS

This overview is based on the existing 3GPP releases 7 and 8 specifications. Release 7 was declared frozen in TSG SA 36 in June 2007. In this release, emergency calls for IMS are supported. Within this context, 3GPP has set the requirements from both the service and regulatory points of view and are described in [5]. The emergency architecture specification is described in [6] and the corresponding protocol requirements and details are discussed in the 3GPP documents [5] and [7]. The emergency support framework for IMS is depicted in Figure 2 and could be summarized in the following points (for more details, we refer to [5]):

- Emergency IMS registration: can be used only to place emergency calls.

- If the User Equipment (UE) has already retrieved its location information, it will include it in the initial request of the emergency call.

- Otherwise the P-CSCF might query for the user location from the access network and refer it in the request. Then forward the request to the Emergency CSCF (E-CSCF).

- Upon receiving the emergency related SIP message by the E-CSCF, in case no location information was provided, the E-CSCF will query the Location Retrieval Function (LRF) for the user location. The LRF may contain or interface with a Routing Determination Function (RDF), ensuring that the E-CSCF will receive the most appropriate Public Safety Answering Point (PSAP) URI, e.g. Police call taker. Then the emergency SIP message is forwarded further to this PSAP. The emergency support for IMS also provides a general mechanism to deal with the callback issue.

- Depending on the user privacy, the PSAP client can extract the location from the SIP message or can get the updated one using the Le interface.

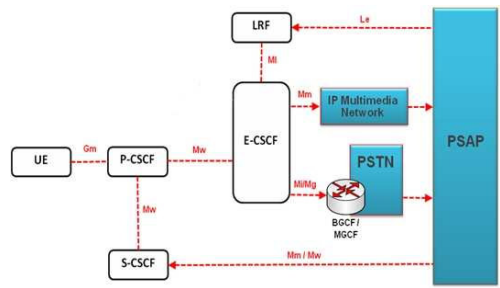

Figure 2: The emergency support framework for IMS

## 2.3   Security Issues in Emergency Services

In the literature, the work on security for emergency calls is very scarce. So far, only [8] has discussed the security threats as well as the setting of requirements to deal with them. The attacks and misuse scenarios that the mentioned characteristics can cause, described in [8], could be summarized in the following points:

- A semantic threat is a wrong indication of the emergency location, either by false testimony by the user or by manipulation of the calling device that automatically provides the location

- information (e.g. a GPS device). This can be exploited in different situations. In an emergency situation, an attacker can try to prevent help reaching its destination, by leading help resources to wrong locations.

- Prioritized traffic, especially combined with unauthenticated access is an ideal target for misuse. Fraudsters can try to misuse the system by using the emergency indicator into normal calls, e.g. during New Year celebrations.

- Breaching the emergency indicator system would also allow attacks on the entire emergency system, e.g. a Denial-of-Service (DoS) attack by flooding

- An attack at the mapping service might rend the emergency service not operational or emergency calls might be wrongly routed, with the same effect as described in the first topic. There are basically three possibilities to attack this node: launching a DoS attack on it, gaining access to it or launching a Man-in-the-Middle attack between it and any contact point

- Emergency information is also sensitive information. It has to be ensured that emergency traffic can not be snooped. Attacks can also be conducted in indirect ways. An example of how to disturb the service without actually launching an attack is to broadcast the (wrong) information that the 911 or 112 system is down and unresponsive. People will get interested or even panic and try to reach the emergency system to check if it's true. This may cause an indirect Denial-of-Service attack by hogging all available emergency lines.

RFC 5069 is without doubt a good starting point to implement built-in security mechanisms in the emergency services support for VoIP networks. However, it should be mentioned that this RFC stays a high level work and some of the scenarios described there might fail to occur in special environments like IMS.

## 2.4 Denial-of-Service Attacks and IMS

DoS attacks are a common security threat in the Internet trying to utilise the target's available resources with the aim of rendering the offered service unavailable [16]. These resources might be bandwidth, CPU or memory. Such threats can also occur in the IMS environments, however, some appropriate application-layer attacks will be used.

In previous papers the vulnerabilities that allow dedicated SIP attacks to occur have been listed [16], [18], [19]. Basically these are missing or wrongly applied sender authentication for packets, software errors in SIP implementations or poorly designed implementations that allow resource depletion to occur.

When talking about a DoS attack, one generally means flooding attacks that overwhelm the victim's resources. In our case, flooding can be achieved with different SIP messages (emergency INVITE, emergency REGISTER, etc.) and the attack can be launched from a single source or from multiple sources. The latter is called a Distributed Denial-of- Service (DDoS) attack where the attacker employs a large number of (usually unaware) computers with different IP addresses to generate a higher-bandwidth stream of messages than would be possible from one single machine.

An Emergency Registration attack can be generated as a wellformed REGISTER request is processed and generates a transaction at least on the P-CSCF and I-CSCF, because the P-CSCF does not check if the user that registeres really exists. Only after interrogating the HSS about the user identity can the I-CSCF decline the registration. In this way registration for spoofed users could be used to attack the P-CSCF of the attached network or the I-CSCF and HSS of the home network. Another vulnerabilty could be when the network has a policy of accepting anonymous emergency calls. This means the requests that include the emergency indicator and belong to a (potential) dialog will be forwarded by the P-CSCF to the E-CSCF without verifying user authentication. Based on this, a DoS attack could target the P-CSCF and the components handling the emergency call, e.g. the E-CSCF. In case the malicious emergency INVITE contains location information the attack could include also the mapping service and even the PSAP, by occupying all the call-takers. The mapping service could be enhanced with a method to verify the validity of the location information for a authenticated user, based on the associated IP, but for an anonymous emergency call this will not be possible, instead a semantic checking can be carried out.

Furthermore, attacks where source IP addresses in packets are spoofed to escape detection can be considered to be a kind of distributed attack. This means also that although the semantic check might eliminate the attacks with location information coming from a valid IP, it will not prevent the processing of the INVITE request on the P-CSCF, E-CSCF and LRF in attacks with spoofed IPs.

### 2.5 Firewalls

A firewall is a network security component deployed at the border of the network. Incoming and possibly outgoing traffic is examined at the firewall according to security policies. The security policies define what kind of traffic will be forwarded or dropped at the firewall. Stallings [20] defines three common firewall types, depending on the layer they are operating: packet-filters (network-layer), application-layer gateways (application layer) or circuit-level gateways (application layer, session level). The essential part of a firewall is the implementation of its security-policy handling. Generally, security policies are given as firewall rules in the form of "if condition then action". Here, condition is related to the packet currently being examined and could be a comparison to an IP address, port or certain content in the message payload, for example. Action consists of either accepting or discarding the examined packet. All defined firewall rules form its rule set. Rules can be static, e.g. the network security operator has defined these rules manually, or updated dynamically by a firewall controller - commonly an Intrusion Detection System (IDS). For each passing packet, the firewall searches the rule set for applying rules (the condition matches the examined packet), and performs the defined action on the packet.

### 2.6 Greylisting

Greylisting [15] is a complementary mechanism to white and black lists used in email spam prevention. When a message is received by an email server from a sender that is not listed on a white or a black list then the message is rejected temporarily. Senders that implement the Simple Mail Transfer Protocol (SMTP) [21] specification would hence correctly retry sending the message later. The re-transmitted message would then be accepted by the server and forwarded to the client. Thus greylisting is based on the assumption that SPAM software is rather simple and is optimised to send a lot of messages but does not care about re-transmissions. This way, messages from legal users would never be dropped unnecessarily and would always be forwarded to the receivers, albeit a bit delayed, while messages from spammers are dropped with very little effort.

## 3 Related Works

As mentioned in section 2.3, RFC 5069 discusses the security threats and requirements for emergency calls making, however, no concrete solution is provided. In [43], we tried to concretize more this RFC by showing some practical attacks (related to the missuse of the emergency identifier) that can occur against the IMS-based emergency framework as well as the corresponding solutions. The current paper is the continuation of such activities where another security issue (flooding attacks) is considered. It is worth to mention that even the solution suggested in this paper is bound to the IMS-based emergency framework, it can be used for any SIP service regardless the infrastructure on top of which it is deployed (IMS or not).

Our work addresses the Denial-of-Service vulnerability on the P-CSCF component as it is the entry point to the IMS network and in particular to the emergency framework.

DoS handling strategies have been discussed in literature in various forms. As there are both multiple and different types of DoS attacks, there is no unique solution that is able to cover all types of attacks [16]. Different approaches have therefore been proposed. Initial approaches for DoS protection have been simple rate-limiting algorithms that allow a limited number of requests per time interval from each sending IP address [23]. These mechanisms are effective for singlesource DoS attacks but fail for highly distributed DDoS attacks. Furthermore, it has been shown in [24] that such mechanisms prove to be a target for a self-inflicted DoS. Several researchers have proposed mechanisms to detect Denial-of-Service attacks using state-machine specifications [12],[25],[9]. Here SIP state machines are modelled for Transactions or Sessions and every SIP message is evaluated within the state machine whether it deviates from the SIP specification ([12], [25]) or depending on which timings occur in the state machine [9]. These mechanisms are helpful against single-source DoS attacks and can also detect DDoS attacks, but without the possibility of mitigating DDoS attacks. Other researchers have developed lightweight statistical algorithms to detect DDoS attacks, e.g. by using Hellinger Distance calculation [14] or calculating cumulative sums [11], [13]. These algorithms can successfully detect DoS and DDoS attacks on SIP proxies, but do not allow any prevention mechanisms. A final group of researchers are focussing on developing combined solutions to detect multiple types of attacks [26], [27],[28], however none of them are able to deliver DDoS mitigation mechanisms.

DDoS mitigation on SIP proxies is a hard problem to resolve. Until now, next to our work we are only aware of one other proposal for handling DDoS attacks on SIP proxies. Nagpal et al. [29] propose a null-authentication mechanism where all unknown requests are challenged with a SIP authentication message that every regular VoIP client will be able to handle but that attackers will generally not be able to process. Our work is based on similar assumptions, however it has the advantage that no additional proxy is necessary to actively generate requests, thus lowering processing overheads. In this paper we do not address further SIP vulnerabilities. Other researchers have proposed mechanisms to handle message tampering (e.g. [30]), fraud and billing attacks (e.g. [31]) or VoIP SPIT (e.g. [32]).

## 4 Proposed Solution for Prevention

Our proposed mechanism works as follows: a firewall with pinholing capabilities is positioned in front of the P-CSCF that needs to be protected, as this entity is the entry point to the IMS network. There is no need to correlate the pinholing information from multiple P-CSCFs in order to exclude the harmless users, because all the traffic of the same non-malicious user will be sent to the same P-CSCF used for registration. Our software solution can have one or muliple monitoring targets as it is build on a fully scalable monitoring architecture. It can be deployed just on one or more machines to scale to higher network loads. The firewall is initially configured to block all incoming requests destined to a P-CSCF, i.e. no pinholes are established. Any arbitrary but regular SIP User Agent (UA) sending a SIP request to a P-CSCF (for example an emergency or non-emergency INVITE or a REGISTER request) will have this request discarded at the firewall. However, afterwards the firewall establishes a pinhole so that further requests with a relation to this UA shall pass the firewall unhindered. Thus, further communication of this UA will not be affected. Because of SIP's defined re-transmission algorithm, it is guaranteed that the UA will automatically re-transmit all SIP messages that have been initially blocked by the pinholing firewall. Consequently, the UA's re-transmission message will pass through the created pinhole to the P-CSCF, thus the communication channel is established. After some time of inactivity (i.e. no traffic passing through this pinhole) the generated pinhole is closed at the firewall. The schematic overview is depicted in Figure 3. A DoS attack on the other hand strives for effective binding of all resources at the SIP server, like the one from the P-CSCF, e.g. through the establishment of as many different transactions as possible (for example by launching a memory depletion attack as described above). Hence, for effective resource depletion the attacker needs to initiate multiple different transactions, which will all be blocked by the access firewall. Only if the attacker UA conforms to the SIP specification and re-sends all previous requests (meaning, it also implements a message timeout detection scheme), will it be able to pass through the pinholing firewall.

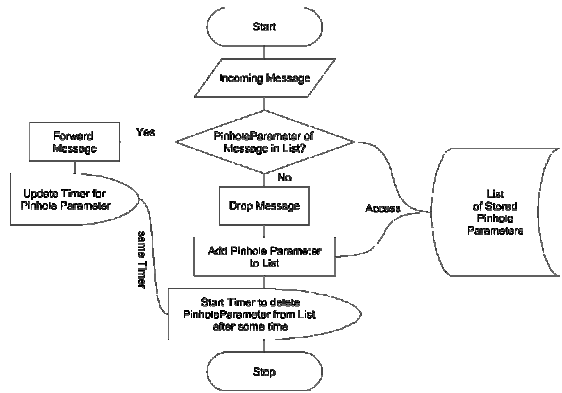

Figure 3: Pinholing process overview

A key component of this mechanism for a successful prevention will be the method for creating pinhole rules at the perimeter firewall, e.g. which parameters should be considered for specifying the pinholes. Here we are considering different possibilities for the pinhole parameter. If we consider a basic attack tool, it is likely that it will generate attack traffic with spoofed source IP addresses. This is necessary, as otherwise the attack will be filtered out by common security solutions. Most commercial SIP security products (e.g. Borderware SIP Assure [33]) use threshold-based prevention mechanisms against DoS. This is achieved by allowing only a limited number of requests from one source within a given time frame. However, such a mechanism is not effective against messages with spoofed IP addresses. For the pinholing algorithm to be effective against such attacks, the pinhole parameter can simply be the source IP address, i.e. every first message from one given IP address will be dropped, while further requests from this same address will pass through (see Figure 4). With this pinhole parameter, prevention is also possible for DDoS attacks with real IP addresses if the attack generation tools does not implement the correct SIP retransmission method. However, with the IP address as the only pinhole parameter, it would be ineffective if attackers were to initiate multiple requests with their real

IP address or with random but fixed spoofed IP addresses. In this case only the first request would be blocked while all further requests could pass unhindered through the newly generated pinhole. To cope with this situation, the pinhole parameter can also be modified to consider the transaction or session ID (i.e. evaluating the relevant Via, or Call-ID header fields and tags in the SIP message) as the pinhole parameter, thus only allowing messages from the same context to pass the firewall.

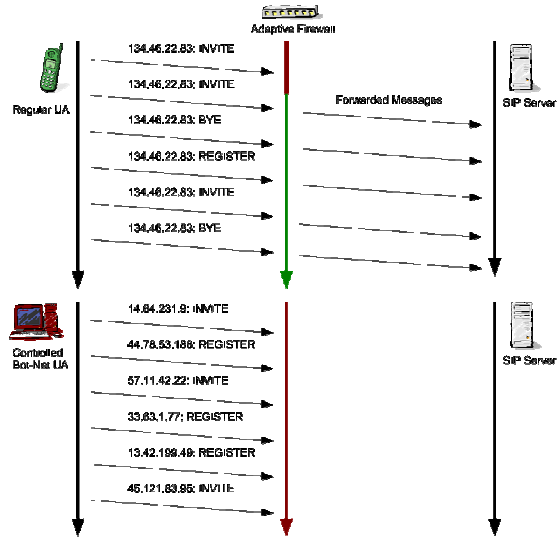

Figure 4: Pinholing overview. The adaptive firewall blocks all new requests. A pinhole is only opened after a request re-transmission and then traffic passes through it unhindered.

It is possible for an intelligent attacker to circumvent this method. However, to achieve this the attack tool needs to be more complex, and thus becomes less effective: instead of using the full given bandwidth capacity to generate different requests, it has to re-send previous messages. As an attacker will not know how many requests have to be sent to finally pass the prevention mechanism, its individual attack power decreases with every re-send: assuming an attacker can generate 1000 attack requests per second, it would only be able to pass 500 individual requests if all of them have to be repeated. This number would further decrease if the attacker had to repeat requests more than two times. The generated pinholes should later be removed from the firewall to free up resources and to speed up firewall performance. Note that for DoS protection it is not absolutely necessary to close the pinholes immediately: an attack can only "hit" an open pinhole by coincidence, and a more sophisticated attack would also pass an already closed pinhole. Also, if pinholes are closed too early, regular users are also affected, as they have to re-open the pinhole. As REGISTER messages are recurring messages sent by the UA, the pinhole should be at least open for the common register refresh time (which defaults to one hour). Note: The pinholing mechanism works for UDP transport, which is the dominant form of SIP transportation and likely to be used in attack scenarios. SIP does not use its own re-transmission feature when sending over a reliable protocol like TCP, here the re-transmission feature of TCP is used instead.

## 5   Implementation and Testbed

For testing, we have implemented a pinholing prototype using the VoIP Defender framework [34]. VoIP Defender is a general SIP security framework allowing rapid prototyping of different SIP security applications. It provides a network sniffer, packet reassembly, a SIP parser and the possibility to add security algorithms as shared libraries. When security incidents occur, notifications can be generated or a firewall can be controlled to be updated with new access policies. VoIP Defender runs on Linux machines and controls an enhanced version of the default Linux firewall iptables. VoIP Defender passively sniffs traffic passing to and from the host that should be protected. Contrary to a Session Border Controller (SBC) it does not operate as a proxy but as a traffic repeater. Thus it is completely invisible to the attacker (except the iptables firewall), which makes it difficult for attackers to launch a direct attack against VoIP Defender itself. For the pinholing mechanism, VoIP Defender captures all traffic to and from a SIP proxy and forwards it to the pinholing algorithm module. The pinhole database from Figure 3 is managed directly there. The VoIP Defender firewall controller is then responsible for installing and removing pinholing rules at the used firewall (see Figure 5). Rules are updated at the firewall as soon as the incidents occur. Note, that while tests have only been conducted with VoIP Defender, the actual pinhole mechanism should also work in other security frameworks like IDS or SBC.

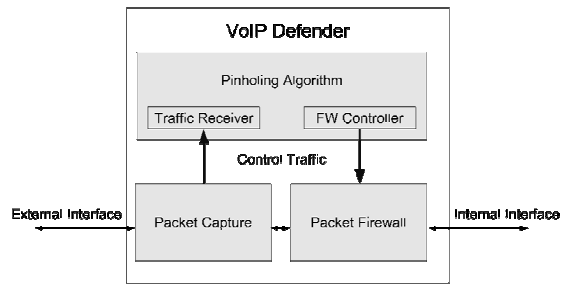

Figure 5: VoIP Defender internal structure

VoIP Defender with the pinholing algorithm is deployed within our security testbed using multiple Dell Xeon servers running the Linux operating system. Incoming traffic and internal connections are routed over Gigabit Ethernet lines and switches. For the attack target we have deployed the Emergency Branch of the Open IMS Core [39] with one P-CSCF. The P-CSCF acts as SIP proxy based on the SIP Express Router (SER) [35]. We use an enhanced version of the SIPp traffic generator tool [36] to create background or "normal" SIP traffic utilising the IMS network. SIPp allows traffic scenarios to be automatically generated in a predefined way. It acts as a common UA to initiate and respond to requests. Our enhancements to SIPp include the possibility to create random events in the scenario and diversifying reply messages using regular expressions. We use this to simulate regular users that call each other randomly. We use an internal developed tool that is able to generate SIP messages with spoofed IP addresses to simulate DoS attacks. This traffic is directed towards the P-CSCF with the aim of disturbing proxy operation. The testbed setup can be seen in Figure 6.

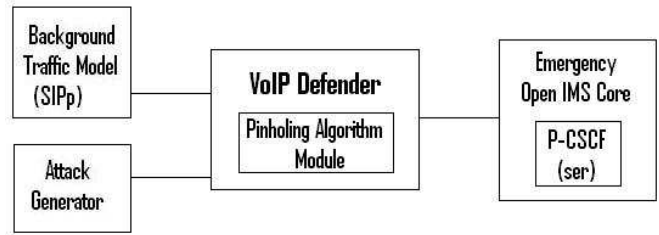

Figure 6: Testbed setup

## 6 Test runs

We created an operation evaluation scenario and a performance evaluation scenario. In the operation evaluation scenario we prove the general operation of the mechanism, whereas in the performance scenario we push the implementation to its limits.

For the operation evaluation scenario, we launch the SIPp background generator from multiple hosts to register different UAs at the IMS network through the P-CSCF and initiate random emergency and non-emergency SIP INVITE requests. Then we start an attack with the attacker tool, by generating similar random requests with spoofed IP addresses starting at a rather low rate of 20 calls / second. During the entire attack the goal is to establish all user-initiated requests from SIPp (i.e. no false positives) and to block all attack-generated requests (i.e. no false negatives) at the firewall. Rules are generated at VoIP Defender's firewall controller in real-time: as soon as a new SIP request is encountered, a new firewall pinholing rule is created and forwarded to the firewall to be installed there. The test shows that all regular users are able to successfully pass the protection solution.

In the next step we measure the delay in processing a regular user's call, i.e. by taking the time between the initial request from the UA and the received answer from the P-CSCF. Without the protection solution established, this delay in the testbed is on average 0.21 s for emergency calls containing the location information and 0.14 for non-emergency calls. With the protection solution established, this delay increases to an average value of 0.71 s and 0.64 respectively. This delay is of course due to the first request being blocked by the firewall, and only the second re-transmission request being able to pass the firewall pinhole. The SIP specification states that the first retransmission message should be generated after $T1$ s, where $T1$ is the calculated RTT or 500 ms, which is exactly what we witness in the test (Note, that SIPp does not seem to do an RTT calculation and just uses the default value of 500 ms for $T1$).

In the second test we measure the performance of the mechanism and its influence on latency. From the previous test we know already the the system works correctly. Hence, it is sufficient to launch the attack alone to generate a high load of requests. We evaluate if the firewall can handle all requests, i.e. dynamically update the firewall table in real-time as requests arrive. In this test, we configure the attack generator to emit 10000 SIP requests at a rate of 500 msg/s, as such a rate would already put a provider system under high stress. Note that the limiting factor here for the SIP proxy of the P-CSCF is mainly processing speed and memory consumption, and not bandwidth usage. All tests are repeated 10 times, to minimise measurement errors. The results from this scenario can be seen in Figure 7. The figure shows that VoIP Defender indeed generates all 10000 firewall pinholes as expected, i.e. there are no false negatives.

However, as more rules are added to the firewall, rule adding latency increases considerably. In a real-time scenario, rules should be effective as soon as new SIP requests are encountered. In the introduced scenario with 10000 SIP requests at a rate of 500 msg/s, ideally after about 20 seconds all rules should have been installed at the firewall. However, the tests show that the last rule is added nearly 3 minutes after the last attack request was generated. This is a considerable delay and would have grave consequences for network traffic. In the theoretical case that only one regular user would contact the P-CSCF just after these 10000 attack messages have been generated, it would take the client three minutes before a contact with the IMS network would have been established. This is however a theoretical value - according to the SIP specification, a calling UA terminates an INVITE request after around 30 s (depending on RTT, and given that the calling human operator would not have given up before). During this time the UA would also have generated ca. 7 re-transmission requests. Evidently, this is not an acceptable use case, and due to the additionally generated re-transmission requests, the load on the defence node would even be increased. Upon examining the setup, we identify the iptables based firewall engine controlled by VoIP Defender as the performance bottleneck. While rules are generated without delay at the firewall controller in VoIP Defender, iptables cannot process these requests in realtime. The standard Linux firewall iptables works considerably well with a small set of static rules, but has known performance problems with an increasing rule set size and rules that are dynamically updated [37, 38, 40]. Within our test runs, there is no great difference in latency performance between different iterations of the test, i.e. derivation from the average case was minimal. This further confirms that the encountered low performance is not due to network problems and indeed results from iptables' rule processing.

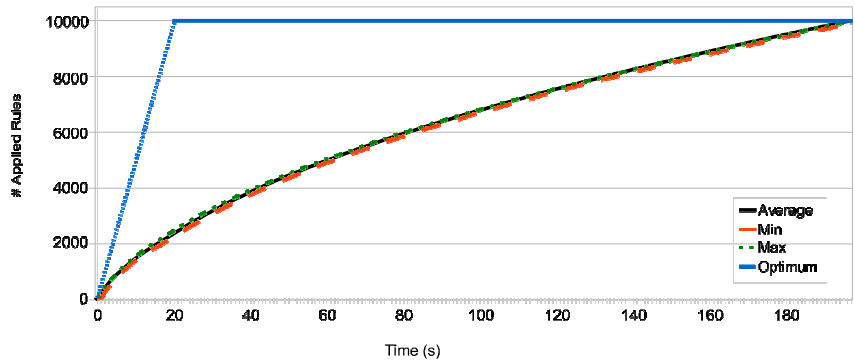

Figure 7: Time to install 10000 rules at the pinholing firewall

## 7 Optimization

The low performance of iptables is mainly due to its costly memory copy operation. As already described, VoIP Defender sends a rule update request in real time, i.e. as soon as the pinhole mechanism decides that a rule update (rule insert / delete) is necessary. However, for each new rule that is modified at the iptables firewall, the whole current firewall rule set has to be copied from kernel-space to user-space, where the rule set is updated. Then, the new rule set is copied back to kernel-space where it is installed in the actual iptables firewall to become effective. Kernel-space to user-space copy operations in Linux are a costly operation. Hence, such a process would be acceptable for static rule entries, however it is not feasible in our case with dynamic rule updates in real time. We therefore propose another solution: instead of updating rules immediately after they are generated by VoIP Defender, they are accumulated at VoIP Defender for a short time, so that a collection of rule update requests can be transferred to the firewall in one control message. These update requests are scheduled periodically, e.g. every second. As a consequence, the iptables firewall can avoid multiple costly memory copy operations, as it can update multiple rule set entries in just one memory copy operation.

This method however has an effect on regular users as the pinholes for regular users are not established immediately after the user's initial request, but only after a scheduled rule update push occurred. However, this effect only affects the signalling path and in effect might cause both a short and acceptable delay at the sender side before the caller hears the waiting tone. We therefore modify the VoIP Defender firewall controller to accumulate individual rules and send out combined update control messages in 1 second intervals. The performance increase with the same scenario is depicted in Figure 8. As can be seen in the figure, the update process is close to the optimum case - the best-case scenario actually matches the optimum case. Even the worst test run is only slightly behind schedule: While 10000 messages are generated within 20 s, all rules are installed at VoIP Defender after just about 21 s, in comparison to the nearly 200 s delay in the original setup.

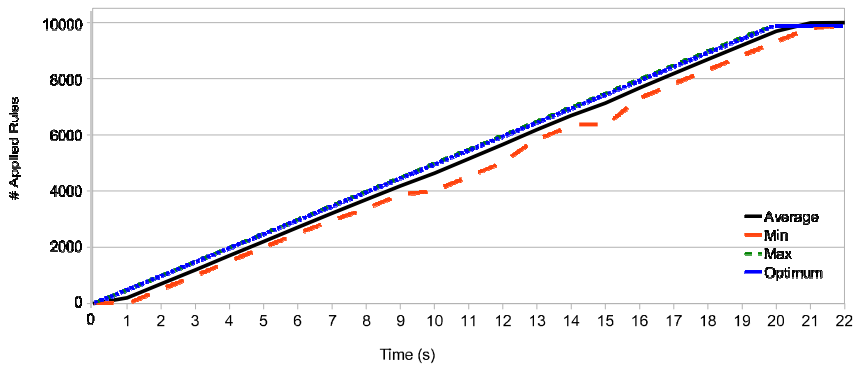

Figure 8: Time to install 10000 rules at the pinholing firewall (optimised version)

As the optimised version works reasonably well, we run another attack with 50000 attack messages. The results are shown in Figure 9.

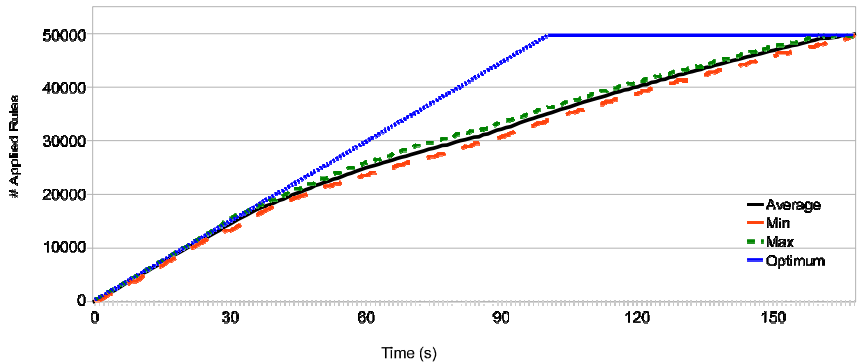

Figure 9: Time to install 50000 rules at the pinholing firewall (optimised version)

With this setup, we can see that iptables can handle up to around 18000 rules in real-time. If further rules are added, latency increases, resulting now in a rule installation delay of 68 s in the worst case. Clearly this test shows that iptables is not the perfect tool to be used as the firewall component, even after applying optimisation strategies. For comparison we summarise the rule-adding speed of the three tests in Table 1. This table shows iptables' rule-adding capabilities at the beginning of each test (without any previous rules established at iptables) and again at the end of each test. The figures are calculated from the lowest performing test run to indicate a worst-case scenario.

Table 1: Worst case rule adding capacities

|  | init. speed | fin. speed |
|---|---|---|
| r/t rule addition, 10000 rules | 191 r/s | 28 r/s |
| 1 s delay addition, 10000 rules | 500 r/s | 433 r/s |
| 1 s delay addition, 50000 rules | 499 r/s | 184 r/s |

While we have shown that the mechanism works as expected, it is evident that iptables cannot cope with the generated traffic load. To circumvent this, one option could be to replace iptables with another firewall solution with better performance (e.g., software iptables replacements like ipset [41] or nf-HIPAC [42] or even a hardware-based solution). On the other hand, other optimisation options exist that could work even with an iptables based solution. One option would be to create the pinhole not after the first request, but after the first re-transmission message is encountered. In this case, the amount of generated firewall rules would be significantly reduced; for the target case of spoofed message attack prevention it is very likely that no firewall rules would be generated at all. On the other hand, regular users would encounter a minimal delay in message processing. Given an arbitrary UA sending an INVITE to the IMS network, its first re-transmission message would be generated after $T1$ s. After this message, the pinhole
would be generated, so that the second re-transmission message after $2 * T1$ s could reach the P-CSCF unhindered by the firewall. Hence, given a responsive proxy the user would experience a delay usually shorter than 2 s, before its request would be processed regularly. Such a delay is not uncommon for a voice scenario, e.g. in mobile call establishments, and thus would likely not irritate the user.

## 8  Future Work

The emergency call delay is bigger than the non-emergency one because it includes also a location parsing and database query for the nearest PSAP URI. In the near future we are considering analysing the performances of the Emergency Branch of Open IMS Core in order to find out what could be the possible optimisations of the message processing during an emergency call.

## 9  Conclusion

In this work we have presented a simple but effective solution to defend special kinds of distributed Denialof- Service attacks on SIP based networks, particularly IMS. The solution adds a marginally longer delay for regular users, but keeps the proxy entity, in this case the P-CSCF, safe from overhead traffic. One major problem we have encountered is the low performance of the used Linux iptables firewall, which is neither capable of handling dynamic rule modification nor scales well beyond 20000 rules in real-time scenarios. While we have shown the general feasibility of our defence mechanism, further stressing it does not seem to be a prudent choice, as long as iptables is still involved. Iptables' limitations show despite our implemented optimisations. Nevertheless, we have proven the effectiveness of the general mechanism, so with a higher performance firewall system, particularly with a hardware-based firewall, it should easily cope with an even higher load. With such a system, pinholing on the transaction or session level instead of just on the IP address level would make the system even more effective. This algorithm can handle a certain type of DoS attack - for effective mitigation, different algorithms should be combined, e.g. utilisation of a lightweight detection scheme (e.g. [13]) to detect DoS traffic, and activate special mitigation schemes like this pinholing solution only if an attack condition has been detected. The pinholing mechanism is therefore an ideal (complementary) companion to common threshold-based protection schemes. A defence solution combining the pinholing mechanism with other DoS protection mechanisms should therefore increase the security of the protected host considerably. We are continuing our research into optimum mitigation schemes under different attack conditions.